\documentstyle[12pt]{article}

\def\noi{\noindent}

\begin{document}
\begin{titlepage}
\rightline{PRA-HEP-92/??}
\vskip 2em
\begin{center}{\Large \bf THE q-BOSON REALIZATIONS OF
THE}\\{\Large \bf QUANTUM GROUP  $U_{q}(sl(n+1,C))$  }\\[6em]
 \v{C} Burd\'{\i}k${}^{\diamond }$,
 L.\v{C}ern\'{y}${}^{*}$
and O. Navr\'{a}til${}^{**}$ \\[2em]
{\sl${}^{\diamond}$
 Nuclear Centre,Faculty of Mathematics and Physics,Charles University,\\
V Ho\-le\-\v{s}o\-vi\v{c}k\'{a}ch 2, 180 00 Prague 8, Czechoslovakia\\
{}$^{*}$
Faculty of Nuclear Sciences and
Physical Engineering,Technical University of Prague,Trojanova 13,
120 00 Prague 2, Czechoslovakia\\
and\\
{}$^{**}$
 Institute of Geotechnics, Czechoslovak Academy of Sciences,
 V Ho\-le\-\v{s}o\-vi\v{c}k\'{a}ch 41, 182 09 Prague 8, Czechoslovakia}
 \\[6em]
\end{center}
\begin{abstract}
 We give explicit expression of recurrency formulae
 of canonical realization for quantum enveloping algebras
 $U_{q}(sl(n+1,C))$. In these formulas the generators
 of the algebra  $U_{q}(sl(n+1,C))$ are
 expressed by means of n-canonical q-boson pairs one
 auxiliary representation of the algebra  $U_{q}(gl(n,C))$.
\end{abstract}

\vskip 4cm

\noi PRA-HEP-92/13

\noi September 1992

\noi \hrule
\vskip.2cm
\hbox{\vbox{\hbox{{\small{\it email
 address:{\small Bitnet=(VSETIN AT CSPUNI12)}}}}}}

\end{titlepage}
\newpage
\setcounter{page}{1}
In a recent paper (Fu and Ge 1992) the authors gave a general method to
construct the q-boson realizations of quantum algebras from their Verma
representations.The method was illustrated on two examples of algebras
$U_{q}(sl(2,C))$ and $U_{q}(sl(3,C))$. In the case of Lie algebras
 this method was formulated
in (Burd\'{\i}k 1985) and some generalization for $U_{q}(sl(2,C))$ in
(Burd\'{\i}k a Navr\'{a}til 1990).

In this letter we are devoted to studying explicitly the general case
  $U_{q}(sl(n+1,C))$. Because it is difficult to write down the explicit
expression of its Verma representation we use the recurrency from
  $U_{q}(gl(n+1,C))$ to  $U_{q}(gl(n,C))$.

 In final formulae the generators of the algebra  $U_{q}(sl(n+1,C))$
are expressed by means of n-canonical boson pairs
and auxiliary representation of the algebra  $U_{q}(gl(n,C))$ . The
pure q-boson realizations we can obtain than after recurrency.

 Very similar formulae were used in our paper (Burd\'{\i}k,Havl\'{\i}\v{c}ek
 and Van\v{c}ura
1992) for construction of irreducible highest weight representations
of quantum groups   $U_{q}(gl(n+1,C))$ .

 Now it is clear (Fu and Ge 1992) that these formulas we can used for
construction of parametrized cyclic representations starting from
cyclic representations of q-deformed Weyl algebra. The irreducibilty
of these representations we study and the results will be published.

 We believe that our recurrency method can be used to construct q-boson
realizations for deformations of other semi-simple Lie algebras as well,and
some positive indication for deformation of $B_n$ and $D_n$ have already
been obtained.

\vspace{10pt}

\begin{flushleft}
 The q-Weyl algebras are defined as  associative algebras $W^q_2$ over
$C$ generated by $b^+,b,$ and $q^{\pm N} $ satisfying (Hayshi 1990)

 $$b b^+ - q^{\mp} b^+ b = q^{\pm N}, q^N q^{-N} = q^{-N} q^{N} =1,$$

 $$ q^N b^{\pm} q^{-N} = q^{\pm} b^{\pm} , (b^-=b),$$

which degenerates to the usual W algebras in the limit  $q \rightarrow 1$.

The n-pairs Weyl algebra we obtain as

 $$W_{2n}^q\;=\;W_2^q \otimes W_2^q \otimes ... \otimes W_2^q\;\;\;\;n-times$$

and the different pairs commute.

 The quantum group  $U_{q}(sl(n+1,C))$
  is defined by the generators $k_i,k_i^{-1},e_i,\;$ and $\; f_i $ for
  $i=1,...,n $,and the relations

 $$ k_ik_i^{-1}=  k_i^{-1}k_i =1 \; ,\; k_ik_j=k_jk_i, $$

 $$ k_ie_jk_i^{-1}=  q^{a_{ij}}e_j  \; ,\;  k_if_jk_i^{-1}=  q^{-a_{ij}}f_j,$$

\begin{equation}
  [e_i,f_j] = {{\delta}_{ij}}{{k_i-k_i^{-1}} \over {q-q^{-1}}},
\end{equation}

 $$[e_i,e_j] =  [f_i,f_j] = 0  \; for  \; {\vert i-j \vert} \geq 2,$$

 $$ e_i^2e_{i\pm 1}-(q+q^{-1})e_ie_{i\pm 1}e_i+ e_{i\pm 1}e_i^2 =0,$$

 $$ f_i^2f_{i\pm 1}-(q+q^{-1})f_if_{i\pm 1}f_i+ f_{i\pm 1}f_i^2 =0,$$

 where $(a_{ij})_{i,j=1,...,n}$ is the Cartan matrix of
  $U_{q}(sl(n+1,C))$,i.e.
$a_{ii}=2,a_{i\pm 1,i}=-1$  and $\; a_{ij}=0 \;$ for
$\vert i-j \vert \geq 2$.

 The generators $e_i,f_i$ correspond to the simple roots. According to
(Rosso 1988) and (Burroughs 1990) we introduce the generators

       $$ X_{n}\;=\;e_{n} $$

 and recurrently

\begin{equation}
        X_r\;=\;e_rX_{r+1}-qX_{r+1}e_r\;\; for \;\; r=1,...,n-1.
 \end{equation}

 The generators $ k_i,k_i^{-1},e_i,f_i \; for \; i=1,2,...,n-1 $ and
 $k_n$ form
a subalgebra in  $U_{q}(sl(n+1,C))$ and evidently it is  $U_{q}(gl(n,C))$.
We add to this subalgebra the generator $f_n$ and we obtain again the
subalgebra which we will denote $A$.

There exists a very simply representation $\phi$ of  $A$  in
$U_{q}(gl(n,C))$

	$$\phi(z)y\;=\;z.y\;\; \phi(f_{n})\;y=\;0 \; for
 \; any \; \; z,y \in U_{q}(gl(n,C))$$

It is a left regular representation.
Because  $U_{q}(sl(n+1,C)) (A \otimes z- 1 \otimes \phi(A)z)\;\;\;z
 \in U_{q}(gl(n,C))$
is an invariant subspace of
the left regular of $U_{q}(sl(n+1,C))$ we can define a generalized
Verma representation $\varrho$ as a factor representation of
the left regular of $U_{q}(sl(n+1,C))$ with respect to this subspace.

The representation space of the representation $\varrho$ is given by

 $$ V({\lambda}_{n})\;=\;{(X_{n})}^{m_{n}}
 {(X_{n-1})}^{m_{n-1}}... {(X_{1})}^{m_{1}} \;  \otimes \; U_{q}(gl(n,C))$$

We will denote

 $$ \vert m \rangle \otimes w\;=\;
\vert m_{n},m_{n-1},...,m_{1} \rangle \otimes w =\;
  {(X_{n})}^{m_{n}}{(X_{n-1})}^{m_{n-1}}... {(X_{1})}^{m_{1}} \;
 \otimes \; w $$

 where $w \in U_{q}(gl(n,C))$
and define the representation $\Gamma$ of $W_{2(n)}$ on $V({\lambda}_{n})$ by

 $$\Gamma(b_i^{+}) \vert m \rangle \otimes w \;=
 \; \vert m+1_i \rangle \otimes w\;=\;
 \vert m_{n},...,m_i+1,...,m_1 \rangle \otimes w,$$

 \begin{equation}
 \Gamma(b_i) \vert m \rangle \otimes w=
 \; [m_i]_q \vert m-1_i \rangle\ \otimes w=[m_i]_q\;
 \vert m_{n},...,m_i-1,...,m_1 \rangle \otimes w,
 \end{equation}

 $$\Gamma(q^{N_i}) \vert m \rangle \otimes w \;=
 \; q^{m_i} \; \vert m \rangle \otimes w, $$

$$ \Gamma([N_i+\alpha]) \vert m \rangle \otimes w\;=
 \; [m_i+\alpha] \; \vert m \rangle \otimes w. $$

 Now for explicit construction we will need the commutation relations $X_j$
with $e_i,f_i$ and $k_i$. Starting from here we will take $r\;\;<n\;\;$.

\vspace{10pt}

 Evidently

\begin{equation}
 e_rX_s\;\;=\;\;X_se_r \;\;\; for \;\; r \;\;< \;\; s-1
\end{equation}

because in $X_s$ are included only $e_t$ for $t>s+1$ which commute with $e_r$.

 For further calculation  the following lemma
 will be useful.
\vspace{16pt}

 { \bf Lemma 1} For $r < n$ it is valid

 $$e_r^2X_{r+1}-(q+q^{-1})e_rX_{r+1}e_r+ X_{r+1}e_r^2 =0.$$
\vspace{10pt}

Proof: For $r=n-1$ it is true from the definition of (1). Now by an induction
. We suppose it is valid for $r=k+1$ and we are calculating

 $$ e_k^2X_{k+1}-(q+q^{-1})e_kX_{k+1}e_k+ X_{k+1}e_k^2 $$

 from a definition $X_{k+1}$ we obtain

 $$=\;\; e_k^2(e_{k+1}X_{k+2}-qX_{k+2}e_{k+1})-(q+q^{-1})e_k
 (e_{k+1}X_{k+2}-qX_{k+2}e_{k+1})e_k+$$

$$+(e_{k+1}X_{k+2}- q X_{k+2}e_{k+1})e_k^2 $$

 from the (4) $e_r$ commute with $X_{r+2}$ and we have

 $$=\;\; [e_k^2X_{k+1}-(q+q^{-1})e_kX_{k+1}e_k+ X_{k+1}e_k^2]X_{r+2}-$$

 $$-q[e_k^2X_{k+1}-(q+q^{-1})e_kX_{k+1}e_k+ X_{k+1}e_k^2]=\;\;0$$

if we use the induction condition.
\rule{.2cm}{.2cm}
\vspace{16pt}

Simply from lemma 1 we obtain

\begin{equation}
  e_rX_r\;=\; e_r^2X_{r+1}-qe_rX_{r+1}e_r\;=\;
 q^{-1}(e_rX_{r+1}-qX_{r+1}e_r)e_r\;=\;q^{-1}X_re_r
 \end{equation}

 for $r<n$.

 Samilarly it is possible to prove that

\begin{equation}
  e_rX_s\;\;=\;\;X_se_r \;\;\; for\;\;\; r+1>s.
\end{equation}

We will continue the calculation of commutation between $f_r$ and $X_s$.

{}From the definition (4) and the commutation relations (1) we have

\begin{equation}
 f_rX_s\;=\;X_sf_r \;\;\;for \;\; s>r.
\end{equation}

If $r=s$ the calculation is more complicated than using
the definition and the above relations gives

$$f_rX_r\;=\; f_r(e_rX_{r+1}-qX_{r+1}e_r)\;=\;(e_rX_{r+1}-qX_{r+1}e_r)f_r-$$

$${-[(k_r-k_r^{-1})X_{r+1}-qX_{r+1}(k_r-k_r^{-1})] \over (q-q^{-1})}=$$

$$=\;\;X_rf_r-X_{r+1}{(q^{-1}k_r-qk^{-1}_r-qk_r+qk_r^{-1})
 \over  (q-q^{-1})}$$

and finally we obtain

\begin{equation}
f_rX_r\;\;=\;\;X_rf_r\;\;+\;\;X_{r+1}k_r
\end{equation}.

In the last case $s<r$  $e_s$ and $f_r$  commute and if we use the definition
of $X_s$ we obtain

$$f_rX_s\;=\;X_sf_r+e_s[f_r,X_{s+1}]-q[f_r,X_{s+1}]e_s$$.

If we put $s=r-1$ we have

$$f_rX_{r-1}\;=\;X_{r-1}f_r+e_{r-1}X_{r+1}k_r-qX_{r+1}k_re_{r-1}=$$

$$X_{r-1}f_r+X_{r+1}(e_{r-1}k_r-qk_re_{r-1})\;\; = \;\; X_{r-1}f_r$$.

By a simple induction we prove

\begin{equation}
f_rX_s\;\;=\;\;X_sf_r\;\;for \;\;s<r
\end{equation}.

Now we have the all commutation relation which we need for the explicit
construction of representations $\varrho$.
The next lemma give the explicit form of the commutation relations
 $e_i,f_i$ and $k_i$ with ${X_j}^{m_j}$
\vspace{16pt}

 { \bf Lemma 2} For $r < n$ it is valid

$$e_rX_r^{m_r}\;=\;q^{-m_r}X_r^{m_r}e_r,$$

$$e_rX_s^{m_s}\;=\;X_s^{m_s}e_r\;\;\; for \;\; r\; \not = \;\;s-1,s,$$

$$e_{r}X_{r+1}^{m_{r+1}}\;=\;[m_{r+1}]_qX_{r+1}^{m_{r+1}-1}X_{r}
+q^{m_{r+1}}X_{r+1}^{m_{r+1}}e_{r},$$

\begin{equation}
f_rX_s^{m_s}\;=\;X_s^{m_s}f_r\;\;\; for \;\; r\; \not = \;\;s,
\end{equation}

$$f_rX_r^{m_r}\;=\;X_r^{m_r}f_r\;+[m_{r}]_{q}X_{r+1}X_{r}^{m_r-1}k_r,$$

$$k_rX_{r+1}^{m_{r+1}}\;=q^{-m_{r+1}}X_{r+1}^{m_{r+1}}k_r\;,
k_rX_{r}^{m_{r}}\;=q^{m_{r}}X_{r}^{m_{r}}k_r\;,$$

$$k_rX_{s}^{m_{s}}\;=X_{s}^{m_{s}}k_r\;,  for s \not= r,r+1.$$

The special cases are $f_{n}$ and $k_n$. In these cases we define

$$Y_{n-1}=e_{n-1}\;\;\;$$

and  recurrently

$$\;\;Y_k=e_kY_{k+1}-qY_{k+1}e_k$$

for $k<n-1$ and it is valid

$$f_{n}X_{n}^{m_{n}}\;=\;X_{n}^{m_{n}}f_{n}\;-{{[m_{n}]}_{q}
 \over (q-q^{-1})}X_{n}^{m_{n}-1}[q^{m_{n}-1}k_{n}
 -q^{-m_{n}+1}k_{n}^{-1}]$$

$$f_{n}X_{r}^{m_{r}}\;=\;X_{r}^{m_{r}}f_{n}\;-q^{-m_r+1}{[m_{r}]}_{q}
X_{r}^{m_{r}-1}k_{n}^{-1}Y_r$$

$$k_nX_{n}^{m_{n}}\;=q^{2m_{n}}X_{n}^{m_{n}}k_n\;,
k_nX_{r}^{m_{r}}\;=q^{m_{r}}X_{r}^{m_{r}}k_n\;,$$

Proof: By using the relations (4),(5),(6),(7),(8),(9) and an induction. The
relations of $k_r$ directly from the definition (1).
\rule{.2cm}{.2cm}
\vspace{16pt}

 Using the relations (10) of lemma 1 we obtain the explicit form of the
representation $\varrho$

 $$\varrho(e_r) \vert m \rangle \otimes w \;\;=\;\;{[m_{r+1}]}_q
 \vert m-1_{r+1}+1_r \rangle \otimes w+
 q^{m_{r+1}-m_r}\vert m \rangle \otimes \phi(e_r)w,$$

 $$\varrho(f_r) \vert m \rangle \otimes w \;\;=\;\;{[m_{r}]}_q
 \vert m+1_{r+1}-1_r \rangle \otimes \phi(k_r)w+
 \vert m \rangle \otimes \phi(f_r)w,$$

 $$\varrho(k_r) \vert m \rangle \otimes w \;\;=\;\;
 q^{m_{r}-m_{r+1}}\vert m \rangle \otimes \phi(k_r)w,$$

\begin{equation}
 \varrho(e_{n}) \vert m \rangle \otimes w \;\;=\;\;
 \vert m +1_{n}\rangle \otimes w,
\end{equation}

 $$\varrho(f_{n}) \vert m \rangle \otimes w \;\;=-\;\;{{[m_{n}]}_q \over
 (q-q^{-1})}
 \vert m-1_{n} \rangle \otimes  (q^{-1+{\sum}_{i=1}^{n}m_i}\phi(k_{n})-
 q^{1-{\sum}_{i=1}^{n}m_i}\phi(k_{n}^{-1}))w- $$

 $$-{\sum_{k=1}^{n}}[m_k]_{q}q^{1-{\sum}_{i=1}^{k}m_i}
 \vert m-1_k \rangle \otimes \phi(k_{n+1}^{-1}Y_r)w,$$

 $$\varrho(k_{n}) \vert m \rangle \otimes w \;\;=\;\;
 q^{m_{n}+{\sum}_{i=1}^{n}m_i} \vert m \rangle \otimes \phi(k_{n})w,$$

{}From the explicit form of the representation $\varrho$ we can see
that it is possible to rewrite this representation using the
representation $\Gamma$ (3). This representation $\Gamma$
and representation $\phi$ are faithful representations and we can
formulate the following theorem.
\vspace{16pt}

{\bf Theorem} The mapping $\tau$ defined by formulae

$${\tau}(e_r)\;\;=\;\;q^{N_{r+1}-N_{r}}e_r\;\;+\;\;b_r^+b_{r+1}$$

$${\tau}(f_r)\;\;=\;\;f_r\;\;+\;\;b_{r+1}^+b_rk_r$$

$${\tau}(k_r)\;\;=\;\;q^{N_{r}-N_{r+1}}K_{r}$$

for $r<n$
\begin{equation}
{\tau}(e_{n})\;\;=\;\;b_{n}^{+}
\end{equation}

$${\tau}(k_{n})\;\;=\;\;q^{{N_n}+{\sum}_{i=1}^{n}N_i}k_n,$$

$${\tau}(f_{n})\;\;=\;-{
{q^{{\sum}_{i=1}^{n}N_i}k_n
 -q^{-{\sum}_{i=1}^{n}N_i}k_n^{-1}} \over {(q-q^{-1})}
 }b_n
-{\sum}_{k=1}^{n-1}{q^{-{\sum}_{i=1}^{k}N_i}}k_n^{-1}Y_kb_k$$

is a homomorphism from  $U_{q}(sl(n+1,C))$ to $W_{2(n)}^{q} \otimes
 U_{q}(gl(n,C))$ .
\vspace{16pt}

In this letter we have presented some simple generalization of the
construction (Fu and Ge 1992). The realizations of $U_{q}(sl(n+1,C))$
are in q-boson pairs and in the generators of the subalgebra $U_{q}(gl(n,C))$.
For using our formulae (12) recurrently to obtain the pure
q-boson realizations it is also needed to have an operator ${\tau}(k_{n+1})$.
Evidently from our construction it is possible reformulate for
$U_{q}(gl(n+1,C))$ and after calculation we obtain

$${\tau}(k_{n+1})\;\;=\;\;q^{{\lambda}_{n+1}-{\sum}_{i=1}^{n}N_i}$$

The pure q-boson realizations are a starting point (see Fu and Ge 1992) for
a construction of the cyclic representations in the root of unity. The
properties of these representations we will study in a forthcoming paper.
\vspace{10pt}

The authors are grateful to the members of the seminar of the quantum groups
in Prague especially Dr M.Havl\'{\i}\v{c}ek ,for useful discuccions.
\vspace{16pt}

\centerline{\large \bf References}
\vspace{16pt}

\begin{description}
\item[Biederharn L C    1989]  J.Phys.A:Math.Gen. 22    p.L873
\item[Burd\'{\i}k \v{C} Havl\'{i}\v{c}ek M and Van\v{c}ura T 1992]
 Commun.Math.Phys. in print
\item[Burd\'{\i}k \v{C} and Navr\'{a}til O 1990]  J.Phys.A:Math.Gen. 23 pL1205
\item[Burd\'{\i}k \v{C} 1985]  J.Phys.A:Math.Gen. 18    p.3101
\item[Burroughs N 1990] Commun.Math.Phys.127 p.109
\item[Hayashi T 1990] Commun.Math.Phys.127 p.129
\item[Hong-Chen Fu and Mo-Lin Ge 1992] J.Math.Phys. 33 p.427
\item[Rosso M 1988] Commun.Math.Phys.117 p.581
\item[Rosso M 1989] Commun.Math.Phys.124 p.307
\end{description}
\end{flushleft}
\end{document}